\documentclass[aps,prb,twocolumn,superscriptaddress,reprint]{revtex4-1}
\usepackage{graphicx, color}
\usepackage{amsmath, amssymb, amsfonts, mathrsfs}
\usepackage{times}
\usepackage{epstopdf,epsfig}
\def\beq{\begin{equation}}
\def\eeq{\end{equation}}
\def\bwt{\begin{widetext}}
\def\ewt{\end{widetext}}
\def\bsp{\begin{split}}
\def\esp{\end{split}} 
\newcommand{\tb}[1]{\textbf{#1}}
\newcommand{\eref}[1]{Eq.~\eqref{#1}} 
\newcommand{\fref}[1]{Fig.~\ref{#1}} 
\newcommand{\PN}{Poisson }
\newcommand{\SR}{Schr\"{o}dinger }
\begin{document}

\title{Free-energy functional of the electronic potential for Schr\"{o}dinger-Poisson theory}
\author{Vikram Jadhao}
\email{vjadhao1@jhu.edu}
\affiliation{Department of Physics and Astronomy, Johns Hopkins University, Baltimore, Maryland 21218, USA}
\author{Kaushik Mitra}
\email{kaushik.umcp@gmail.com}
\noaffiliation
\author{Francisco J. Solis}
\affiliation{School of Mathematical and Natural Sciences, Arizona State University, Glendale, Arizona, 85306, USA}
\author{Monica Olvera de la Cruz}
\affiliation{Department of Materials Science and Engineering, Northwestern University, Evanston, Illinois 60208, USA} 
\date{\today}

\begin{abstract}
In the study of model electronic device systems where electrons are typically under confinement, a key obstacle is the need to iteratively solve the coupled Schr\"odinger-Poisson (SP) equation. It is possible to bypass this obstacle by adopting a variational approach and obtaining the solution of the SP equation by minimizing a functional. Further, using molecular dynamics methods that treat the electronic potential as a dynamical variable, the functional can be minimized on the fly in conjunction with the update of other dynamical degrees of freedom leading to considerable reduction in computational costs. But such approaches require access to a true free-energy functional, one that evaluates to the equilibrium free energy at its minimum. In this paper, we present a variational formulation of the Schr\"odinger-Poisson (SP) theory with the needed free-energy functional of the electronic potential. We apply our formulation to semiconducting 
nanostructures and provide the expression of the free-energy functional for narrow channel quantum wells where the local density approximation yields accurate physics and for the case of wider channels where Thomas-Fermi approximation is valid.
\end{abstract}

\maketitle 

\section{Introduction}
In recent years, tremendous progress has been made in the experimental study of semiconducting nanostructures.  Advanced epitaxial techniques allow systematic control of atomic-scale
features in the fabrication along vertical dimension, while nanoscale lithography
allows the patterning of structures with few-nanometer lateral dimensions\cite{Ihn,Steiner}. Fine tuning of these characteristic length 
scales along lateral and vertical directions allow carrier confinement such that the corresponding energy levels are quantized. Examples include InGaN/GaN quantum wells with three-dimensional electron
confinement\cite{nakamura1996ingan}, carbon nanotubes, in which carriers are confined in two dimensions\cite{tans1998room}, and the inversion or accumulation layer of metal-oxide field effect transistors with quasi-2D carrier confinement\cite{datta2005quantum}.

Regardless of the confinement scheme, however, quantization is affected by a carrier potential that
depends on the details of the particular nanostructure embodiment. It is desirable, therefore,
to have methods for calculating the carrier potential that are applicable to a broad class of nanostructures. Theoretically, such systems have been modeled using the coupled Schr\"{o}dinger-Poisson (SP) equation that gives rise to a self-consistent solution for the carrier potential \cite{datta2005quantum,wettstein2000simulation,kim2000self,john2004schrodinger,assad2000performance}. A conventional approach to obtain this self-consistent solution is to solve the \SR and \PN equations iteratively using the finite-difference method. \cite{ram2004schrodinger,ram2006wavefunction,pacelli1997self,trellakis1997iteration,tan1990self,luscombe1992electron}. 

Finite-difference methods can often get slow and expensive due to the iterative nature of the calculations, which involves going back and forth between the Schr\"{o}dinger and Poisson equations. Therefore, it is useful to explore other ways of arriving at the carrier potential. In this paper, we approach the problem of solving the SP equation via a variational principle. 
Variational formulations often shed new light on the underlying physics of the problem as they appeal to the universal physical principle that the true solution of the problem must minimize the free energy associated with the system. Within a variational approach it is possible to bypass any iterative calculations and arrive directly at the carrier potential via the minimization of an appropriate functional.  Furthermore, if the variational principle is based on a true free-energy functional, that is, a functional that minimizes to reveal the equilibrium free-energy of the system, the SP equation can be solved on-the-fly leading to enormous savings in computational costs. For example, using dynamical optimization procedures such as the Car-Parrinello molecular dynamics \cite{car-parrinello}, it is possible to evaluate the evolving electronic potential profile in a double-gate device in conjunction with updating the gate voltage without explicitly re-solving the SP equation. 

Among the variational principles of SP theory that exist in the literature, none of the formulations are based on a free-energy functional \cite{nier,arriola,carrillo}. For example, the functional derived by Nier does single out the correct potential upon minimization, but its minimum value is negative of the free energy \cite{nier}. In this article, we derive the needed free-energy functional of the potential for the SP theory. We construct a local and a non-local form for the free-energy functional. While the non-local functional is applicable to generic systems, the construction of the local form is contingent upon the availability of an inverse function specific to the system under consideration. On the other hand, we expect the local functional to offer more benefits from a computational viewpoint compared with its non-local counterpart. The derivation of the local functional employs the modified Lagrange multiplier procedure for constructing constrained variational functionals recently introduced by us 
in Ref.~\onlinecite{jso4}. We note that the functionals derived here and the associated variational formulation bear similarity to the variational principles we have developed for electrostatics and Poisson-Boltzmann theory \cite{jso1,jso2,jso3,jso4}.

In Sec.~\ref{sec:sptheory}, we outline the principles of the SP theory. Sec.~\ref{sec:var.principle} provides the variational formulation of the SP theory with a true free-energy functional of the electronic potential. In Sec.~\ref{sec:discussion}, we consider model Al/GaAs quantum wells and derive specific expressions of the local functional for narrow and wide channel quantum wells. 
We close with some final remarks in Sec.~\ref{sec:conclusion}. 

\section{Schr\"{o}dinger-Poisson theory}\label{sec:sptheory}
Consider a gas of electrons confined in a $N$-dimensional space. As mentioned above, some common examples include quantum dots ($N=3$), carbon nanotubes ($N=2$), and metal-insulator-metal arrangements where electrons are under quasi-2D confinement. An important quantity of interest is the carrier potential profile for a given set of external conditions. If the density of electrons is a known function of position then, to a good approximation, the carrier potential can be computed from the Poisson equation. In general, however, the electronic charge density is not known, thus making the potential profile inaccessible. The Schr\"{o}dinger-Poisson (SP) theory suggests a way to proceed in this situation. 

Microscopically, SP theory is equivalent to a self-consistent Hartree theory where the many-body effects of electronic interactions are approximated by a mean field given by the solution to the Poisson equation. 
In other words, within the SP formulation of the problem, the potential term entering an effective one-electron \SR equation is taken to be the same potential that solves the Poisson equation whose source term is the local electron density. Note that the electron density is a function of Fermi-Dirac operator, which in turn depends on the Hamiltonian (and hence the original potential) entering the \SR equation. Therein lies the self-consistent nature of the SP formulation reflected in a set of coupled \SR and \PN equations. 

The \SR equation for the effective one-electron system is given by
\begin{equation}\label{eq:schrodinger}
\left[-\frac{\hbar^2}{2m}\nabla^2 + U(\textbf{r}) + V_{\textrm{ext}}(\textbf{r})\right]\psi_k(\textbf{r}) = \lambda_k\psi_k(\textbf{r}),
\end{equation}
where $\hbar$ is Planck's constant divided by $2\pi$, $m$ is the mass of the electron, and $\psi_k$ and $\lambda_k$ are, respectively, the energy eigenfunctions and eigenvalues. The function $U(\tb{r}) = qV(\tb{r})$ is the electrostatic contribution to the potential energy with  $V(\textbf{r})$ being the effective electrostatic potential field which the electron experiences and $q$ is the electronic charge. The external potential $V_{\textrm{ext}}(\textbf{r})$ is the electronic confinement potential that leads to space quantization and in the rest of the paper will be considered as a known function of $\tb{r}$.
The key assumption in SP theory is that $V(\textbf{r})$ satisfies the Poisson equation:
\begin{equation}\label{eq:poisson}
-\nabla\cdot\left(\frac{\epsilon(\textbf{r})}{4\pi} \nabla V(\textbf{r}) \right) = qn(\textbf{r}),
\end{equation}
where $\epsilon(\tb{r})$ is the permittivity and $n(\textbf{r})$ is the electron density which is given by the equation:
\beq\label{eq:density1}
n(\tb{r}) = g\sum_k \left|\psi_k(\textbf{r})\right|^2 f_k.
\eeq
In \eref{eq:density1}, $f_k = (\exp\beta(\lambda_k - \mu)+1)^{-1}$ is the Fermi-Dirac function that provides the occupation number of the $k^{\textrm{th}}$ state with $\mu$ being the chemical potential and $\beta = 1/k_{B}T$ is the inverse thermal energy. The factor $g$ is introduced to account for degeneracies. For example, in the case of electron spin, $g$ assumes the value 2. 

Clearly, equations \eqref{eq:schrodinger} and \eqref{eq:poisson} are coupled and their solution exhibits a self-consistent character. The self-consistent calculation typically proceeds in the following way. One begins with a choice of $V(\tb{r})$ and solves \eref{eq:schrodinger} to obtain $\lambda_k$ and $\psi_k$. Using these results, $n(\tb{r})$ is evaluated from \eref{eq:density1}. This density is employed in \eref{eq:poisson} and the latter is solved for the potential $V(\tb{r})$. If the newly computed $V(\tb{r})$ is the same as the original starting potential choice (or within some pre-defined small error), the calculation stops, otherwise, the loop continues with the new $V(\tb{r})$ as the potential choice for \eref{eq:schrodinger}. 

While the above procedure of obtaining the potential is common in the literature, we would like to emphasize that, within the framework of the SP approximation, we are not required to solve the \SR equation. To see this, we recognize that the density $n(\tb{r})$ is a thermodynamical quantity and is invariant under basis transformation. In other words, the evaluation of $n(\tb{r})$ does not require the eigenbasis of the Hamiltonian operator; any orthonormal basis can be employed, although it is expected that the eigenbasis would offer the most ease in carrying out the computation. Hence, solving the \SR equation is not necessary to evaluate $n(\tb{r})$ and subsequently compute the electronic potential. This fact is useful to represent the aforementioned self-consistency of the SP theory in the form of a single equation which we show next.

The invariance of the density $n(\tb{r})$ under basis transformation is reflected by 
\beq\label{eq:density}
n(\tb{r}) = \left<\tb{r}|f|\tb{r}\right> 
\eeq
where $f$ is the Fermi-Dirac operator 
$f = (\exp\beta(H - \mu) + 1)^{-1}$
with $H = K + U + V_{\textrm{ext}}$ being the Hamiltonian. Here $K=-\hbar^2(2m)^{-1}\nabla^{2}$ is the kinetic energy
operator. For the sake of brevity, we will often omit writing the operators using complete notation, for example, $\hat{f}$ or $\hat{H}$. We will employ the full notation or emphasize the meaning of the symbols when it becomes necessary to clarify the content. Also, for clarity, we will suppress the explicit notation for the identity operator and instead use 1 to denote both the operator and the number and trust the reader to figure out the appropriate meaning from the context. We observe that the operator $f$ is a function of $V$ making $n(\tb{r})$, in general, a functional of $V(\tb{r})$. Accordingly, we introduce the notation $n[V]$ and define the charge density $\rho$ as $\rho[V] = q n[V]$. Note that $\rho$ is also a function of $\textbf{r}$ and sometimes we employ the notation $\rho_{\textbf{r}}[V]$ to express the multiple dependencies of $\rho$.  
Substituting $n$ from \eref{eq:density} in \eref{eq:poisson} and employing the definition of $\rho$, we obtain
\beq\label{eq:spequation}
\nabla\cdot\left(\frac{\epsilon(\textbf{r})}{4\pi} \nabla V(\textbf{r}) \right) + \rho_{\tb{r}}[V] = 0.
\eeq
Equation \eqref{eq:spequation} is the SP equation. The function $V(\tb{r})$ that satisfies the above equation is the equilibrium potential of the confined electron system. 
The electron density $n(\tb{r})$ can be evaluated by substituting this potential in \eref{eq:density} and  the energy eigenvalues and eigenfunctions for the system are known by employing this potential in \eref{eq:schrodinger}.

\section{Variational Principle for Schr\"{o}dinger-Poisson theory}\label{sec:var.principle}
We will now approach the SP formulation of the confined electron gas problem via a variational principle. Within a variational approach, we find the potential that satisfies the SP equation by minimizing a judiciously constructed functional. In the first part of this section, we provide our functional, in both its local and non-local form, and discuss its behavior at the extremum. The second part shows the derivation of the functional. For the sake of brevity, we will often suppress the dependency of various fields on the position coordinate when writing the functional expressions. We will put back the dependency when the use of explicit position variables becomes necessary to clarify the content.
\subsection{Free-energy functional}
We construct a free-energy functional with the electronic potential as the sole variational field. Our SP functional has the form: 
\begin{eqnarray}\label{eq:spfunctional}
F[V] &=& \textrm{Tr}[f(K+V_{\textrm{ext}})] + \frac{1}{8\pi}\int \epsilon \mathbf{\nabla}V \cdot \mathbf{\nabla}V d\textbf{r} \nonumber\\
&+& \frac{1}{\beta} \textrm{Tr}[(1 - f)\textrm{ln}(1 - f) + f\textrm{ln}(f)] - \mu \textrm{Tr}[f] \\\nonumber
&-& \int W[V] \left( -\mathbf{\nabla} \cdot \frac{\epsilon}{4\pi}  \mathbf{\nabla} V - \rho_{\tb{r}}[V] \right) d \textbf{r},
\end{eqnarray}
where $\textrm{Tr}[\ ]$ denotes the trace operation. The choice of the functional $W[V]$ determines whether our functional is local or non-local. For the non-local form, $W = W_{\textrm{NL}}$, where $W_{\textrm{NL}}$ is given by 
\beq
W_{\textrm{NL}}[V] = \int G_{\tb{r},\tb{r}'} \left( \mathbf{\nabla} \cdot \chi_{\tb{r}'} \mathbf{\nabla} V_{\tb{r}'} + \rho_{\tb{r}'}[V] \right) d\tb{r}',
\eeq
where $\chi$ is the susceptibility defined as $\chi = (\epsilon - 1)/4\pi$ and $G(\tb{r},\tb{r}') = |\tb{r}-\tb{r}'|^{-1}$ is the free-space Green's function. For the local free-energy functional, we have $W = W_{\textrm{L}}$ with 
\beq\label{eq:LMLocal}
W_{\textrm{L}}[V] = \rho^{-1}\left(-\mathbf{\nabla} \cdot \frac{\epsilon}{4\pi} \mathbf{\nabla} V \right),
\eeq
where $\rho^{-1}$ is the inverse function corresponding to the charge density $\rho$. Wherever we need to discuss the functionals separately, we will employ the notation $F_{\textrm{L}}$ and $F_{\textrm{NL}}$ to denote the local and non-local functionals respectively.

In \eref{eq:spfunctional}, the first two terms on the right-hand side correspond to the internal energy of the confined electron system treated within the SP formulation. These terms are the expectation values of the kinetic and potential energies of the system. The third term is the entropy of an electron gas according to the Fermi-Dirac statistics. Together, the first four terms correspond to the free energy $I_{\textrm{o}}[V]$ of the confined electron system. In the last term of \eref{eq:spfunctional}, the terms enclosed in the brackets form the left hand side of the SP equation, \eref{eq:spequation}. Thus, in order to find the potential function $V(\tb{r})$ that minimizes the free energy of confined, interacting electrons subject to the satisfaction of the SP equation, we treat the latter as a constraint and in the spirit of the method of Lagrange multipliers write a functional that has the free energy $I_{\textrm{o}}[V]$ constrained by the SP equation with $W[V]$ playing the role of the Lagrange 
multiplier function. As we show below, the function $W$ determines whether the resulting functional has the desirable properties of convexity and meaningful free energy evaluation at equilibrium. 

Employing a compact notation, we can write our functional as
\beq\label{eq:spfnal.compact}
F[V] = I_{\textrm{o}}[V] - \int W[V] C[V] d\tb{r}
\eeq
where we introduce the notation $C[V]$:
\beq\label{eq:constraint}
C[V] = -\mathbf{\nabla} \cdot \frac{\epsilon}{4\pi}  \mathbf{\nabla} V - \rho_{\tb{r}}[V]
\eeq
for the constraint.
When expressed in the form shown in \eref{eq:spfnal.compact}, our functional is identical to the constrained variational functional studied in Ref.~\onlinecite{jso4} where we provide the formulas for the first and second variations of the functional. These formulas can be employed to check the extremal behavior of $F[V]$. Using the result for the first variation \footnote{See Eq.~(33) of Ref. 20}, we find that the condition for $F[V]$ to be extremum is 
\beq\label{eq:ext.cond}
C[V] = -\mathbf{\nabla} \cdot \frac{\epsilon}{4\pi}  \mathbf{\nabla} V - \rho_{\tb{r}}[V] = 0,
\eeq
which coincides with \eref{eq:spequation}, the SP equation. Thus we obtain the self-consistent solution of the SP equation via extremization of our functional. Further, using the general formula for the second variation \footnote{See Eq.~(41) of Ref. 20}, we obtain the second variation of the local functional $F_{\textrm{L}}[V]$ to be
\begin{eqnarray}\label{eq:secondvar.local}
\delta^{2}F_{\textrm{L}} &=& 
3\int \frac{\epsilon}{4\pi}|\nabla\delta V|^{2}d\tb{r} + \beta q^{2} \int \sum_{k}s_{k}|\psi_{k}|^{2} \delta V^{2} d\tb{r} \nonumber\\
&&+ \frac{2}{\beta q^{2}}\int \frac{(\nabla\cdot\frac{\epsilon}{4\pi}\nabla\delta V)^2}{\sum_{k}s_{k}|\psi_{k}|^{2}}d\tb{r},
\end{eqnarray}
where $s_{k}$ is
\beq
s_{k} = \frac{e^{\beta(\epsilon_{k} - \mu)}}{\left(1 + e^{\beta(\epsilon_{k} - \mu)}\right)^{2}}.
\eeq
Noting that $s_{k} > 0$, it is clear from \eref{eq:secondvar.local} that $\delta^{2}F_{\textrm{L}} > 0$. Our local functional becomes a minimum at its extremum. Carrying out an explicit calculation for the second variation of the non-local functional gives the following result:
\begin{eqnarray}\label{eq:secondvar.nonlocal}
\delta^{2}F_{\textrm{NL}} &=& 
\int \frac{3\epsilon-2}{4\pi}|\nabla\delta V|^{2}d\tb{r} +
3\beta q^{2} \int \sum_{k}s_{k}|\psi_{k}|^{2} \delta V^{2} d\tb{r} \nonumber\\
&&+ 2 \iint \sigma_{\tb{r}} G_{\tb{r},\tb{r}'} \sigma_{\tb{r}'} d\tb{r}d\tb{r}',
\end{eqnarray}
where 
\beq
\sigma(\tb{r}) = \int \left( \frac{\delta C[V_{\tb{r}}]}{\delta V_{\tb{r}'}} + \frac{1}{4\pi}\nabla_{\tb{r}}^{2} \delta_{\tb{r},\tb{r}'} \right) \delta V_{\tb{r}'}d\tb{r}'.
\eeq
By examining each of the three terms in \eref{eq:secondvar.nonlocal}, it is easy to see that $\delta^{2}F_{\textrm{NL}} > 0$. Thus, the non-local functional $F_{\textrm{NL}}$ also becomes a minimum at its extremum.

Finally, we examine the value of our functional at its extremum. Employing the extremum condition, \eref{eq:ext.cond}, in Eq.~\eqref{eq:spfnal.compact}, we find
$
F[V^{*}] = I_{\textrm{o}}[V^{*}],
$
where $V^{*}$ denotes the potential at which $F$ becomes a minimum. In other words, our functional is equal to the free energy of the confined electron system at equilibrium. In summary, we find that $F[V]$ singles out the correct potential upon extremization, evaluates to the true free energy of the electron gas at equilibrium, and is minimized at its extremum. 

It is instructive to compare $F[V]$ with the functional derived by Nier. One can show that Nier's functional is the negative of the functional:
$I[V] = I_{\textrm{o}}[V] - \int V C[V] d\tb{r}$. Comparing $I[V]$ with our functional $F[V]$ as expressed in \eref{eq:spfnal.compact}, we find that the two functionals are identical except for the Lagrange multiplier term employed to enforce the constraint of SP equation. While $F[V]$ employs $W_{\textrm{L}}[V]$ or $W_{\textrm{NL}}[V]$ as the Lagrange multiplier, $I[V]$ uses the function $V$ for the same purpose. One can readily verify that while $I[V]$ produces the correct potential upon extremization and evaluates to the correct free energy, it becomes a maximum at the extremum. The functional derived by Nier is negative of $I[V]$ and so while it does become a minimum at the extremum, its evaluation at the minimum point is not equal to the true free energy of the electronic system and hence it is not a free-energy functional. This comparison highlights the importance of the choosing the Lagrange multiplier in the construction of a free-energy functional. In the next subsection, we will show how we arrived 
at the specific forms of the multipliers $W_{\textrm{L}}$ and $W_{\textrm{NL}}$ that endow our functional with the features of a free-energy functional.

We note that the construction of the local functional is contingent upon the existence and computability of the inverse of $\rho_{\textbf{r}}[V]$. As we will show in Sec.~\ref{sec:discussion}, for several models of device structures, we are able to produce the local functional in the important regime where quantum effects are dominant. On the other hand, the expression for the non-local functional is valid for a generic system. However, from a computational standpoint, we expect the local functional to offer more benefits relative to the non-local functional.

\subsection{Derivation}
We now derive our SP functional in \eref{eq:spfunctional}. 
We begin by writing the free energy of an electron gas as the functional:
\begin{eqnarray}\label{eq:fenergy}
I[\tb{E},f] &=& \textrm{Tr}[f(K+V_{\textrm{ext}})] + \frac{1}{8\pi} \int \epsilon \, \tb{E}\cdot\tb{E} \, d\textbf{r} \\
&&+ \frac{1}{\beta} \textrm{Tr}\left[ \left(1-f\right)\ln\left(1 - f\right) + f\ln f \right] - \mu \textrm{Tr}[f],\notag
\end{eqnarray}
where $\tb{E}$ is the electric field. Note that $f$ in the above functional is as yet an unknown (operator) variable; it will turn out to be the Fermi-Dirac function at equilibrium. Only $\tb{E}$ and $f$ are treated as variable fields, other quantities, such as $V_{\textrm{ext}}$, $\mu$, and $\beta$ are considered as parameter fields. Also note that in writing the above expression, we have partially employed the SP approximation; we have identified a part of the potential energy of the Hamiltonian with the electrostatic energy. The first term on the right-hand side of \eref{eq:fenergy} is the expectation value of the kinetic energy and the external potential. The second term is the potential energy as given by classical electrostatics. Together, these two terms form the internal energy of the system. The third term is the entropic contribution to the free energy and the last term arises from the dependency of the thermodynamic free energy on the total number of particles. 

We now introduce Gauss's law as a constraint to the above functional, thus obtaining
\begin{align}\label{eq:fenergy+law}
I[\tb{E},f,V] &= \textrm{Tr}[f(K+V_{\textrm{ext}})] + \frac{1}{8\pi} \int \epsilon \, \tb{E}\cdot\tb{E} \, d\textbf{r} \notag\\
&\quad+ \frac{1}{\beta} \textrm{Tr}\left[ \left(1-f\right)\ln\left(1 - f\right) + f\ln f \right] - \mu \textrm{Tr}[f] \notag\\
&\quad-\int V\left[ \frac{1}{4\pi}\mathbf{\nabla} \cdot \epsilon \tb{E} - \rho_{\tb{r}}[f] \right] d\tb{r},
\end{align}
where $V$, at this stage, is the Lagrange multiplier function used to enforce the constraint of Gauss's law, and $\rho_{\tb{r}}[f]=q\left<\tb{r}|f|\tb{r}\right>$. We note that $V$ will turn out to the electrostatic potential corresponding to the electric field $\tb{E}$ at equilibrium. The above functional is a function of $\tb{E}$, $f$, and $V$. Moving forward, our goal is to express it as a functional of one variational field, and we want this field to be $V$.

Variation of $I[\tb{E},f,V]$ with respect to $\tb{E}$ leads to 
\beq\label{eq:Eateqm}
\tb{E} = - \mathbf{\nabla} V,
\eeq
that is, we recover Maxwell's second law. The above relation also implies that $V$ is the electrostatic potential. Variation of $I[\tb{E},f,V]$ with respect to $f$ gives
\beq\label{eq:fateqm}
\hat{f} = \frac{1}{e^{\beta\left(\hat{K}+\hat{V}_{\textrm{ext}}+q\hat{V}-\mu\right)}+1}.
\eeq
Thus, we find that the (operator) function $f$ that keeps the functional $I$ of Eq.~\eqref{eq:fenergy+law} stationary has the Fermi-Dirac form with the Hamiltonian $H = K + V_{\textrm{ext}} + qV$. 
Equations~\eqref{eq:Eateqm} and \eqref{eq:fateqm} together imply the that the electrostatic potential is equal to the electronic potential appearing in the Hamiltonian for the effective one-electron system. Variation of $I[\tb{E},f,V]$ with respect to $V$ generates
\beq
\frac{1}{4\pi}\mathbf{\nabla} \cdot \epsilon \tb{E} - \rho_{\tb{r}}[f] = 0,
\eeq
which upon substitution of $\tb{E}$ and $f$, from \eref{eq:Eateqm} and \eref{eq:fateqm} respectively, gives
\beq\label{eq:Vateqm}
-\frac{1}{4\pi}\mathbf{\nabla} \cdot \epsilon \mathbf{\nabla} V - \rho_{\tb{r}}[V] = 0.
\eeq
The above equation can be rearranged into the following recursive relation:
\beq\label{eq:recursive}
V = \int G_{\tb{r},\tb{r}'} \left( \mathbf{\nabla} \cdot \chi_{\tb{r}'} \mathbf{\nabla} V_{\tb{r}'} + \rho_{\tb{r}'}[V] \right) d\tb{r}'.
\eeq
Substituting $\tb{E}$, $f$, and $V$ from Eqs.~\eqref{eq:Eateqm}, \eqref{eq:fateqm}, and \eqref{eq:recursive} respectively, back in \eref{eq:fenergy+law} leads to the non-local free-energy functional $F_{\textrm{NL}}[V]$.

To derive the local free-energy functional we first note that from Eqs.~\eqref{eq:Eateqm} and \eqref{eq:fateqm}, we can eliminate $\tb{E}$ and $f$ from Eq.~\eqref{eq:fenergy+law} in favor of $V$. Carrying out these eliminations, we obtain the reduced functional
\beq\label{eq:I}
I[V] = I_{\textrm{o}}[V] - \int V C[V] d\tb{r},
\eeq
where $I_{o}$ is the unconstrained functional defined earlier and $C[V]$ is the function (constraint) given by \eref{eq:constraint}. Recall that $I[V]$ is negative of the Nier functional and it is not a free-energy functional. 
Using the modified Lagrange multiplier method introduced in Ref.~\onlinecite{jso4}, we transform this functional into the local free-energy functional $F_{\textrm{L}}$. Following this method, the modified reduced functional
\beq\label{eq:deriv.local}
F[V] = I_{o}[V] - \int h^{-1}[C(V) + h(V)] C[V] d\tb{r},
\eeq 
becomes a free-energy functional if the function $h$ satisfies the following inequality:
\beq
-\int \delta C[V^{*}]\delta V d\tb{r} - 2\int \delta_{h} h^{-1}[h(V^{*})](\delta C[V^{*}])^{2} d\tb{r} > 0,
\eeq
where $h^{-1}$ is the inverse of $h$. We find that choosing 
\beq
h(V) = \rho[f] \equiv \rho[V],
\eeq 
makes the above inequality true. Substituting $h(V)$ from the above equation and its inverse $h^{-1} (= \rho^{-1})$ back in \eref{eq:deriv.local}, and expanding $C[V]$ using \eref{eq:constraint}, we obtain the local free-energy functional $F_{\textrm{L}}[V]$.

\section{Discussion}\label{sec:discussion}
We now focus our attention to physical systems that exhibit quantum confinement of electrons and can be described using SP theory. In this section, we investigate GaAs heterostructures and obtain the expression for the local functional $F_{\textrm{L}}$ in the limit when the local density approximation (LDA) is valid and in the case where Thomas-Fermi (TF) approximation can be applied. In such systems, the true many-body interaction of the electrons is often approximated by a mean-field Hartree potential given by the solution to the Poisson equation \cite{dassarma1984,dassarma1997,hansen1999,nagraja1997}. Furthermore, the effect of the exchange-correlation part of the electron interaction potential (the Fock term) is given by a density-dependent term that adds to the  mean-field Hartree potential \cite{lundqvist1971}. The resulting modified SP equation is thereafter solved iteratively using LDA, where the net effect of the exchange-correlation potential terms is only to renormalize the (density-dependent) 
chemical potential at every step of the iteration. 

\begin{figure}
\centerline{\includegraphics[scale=0.65]{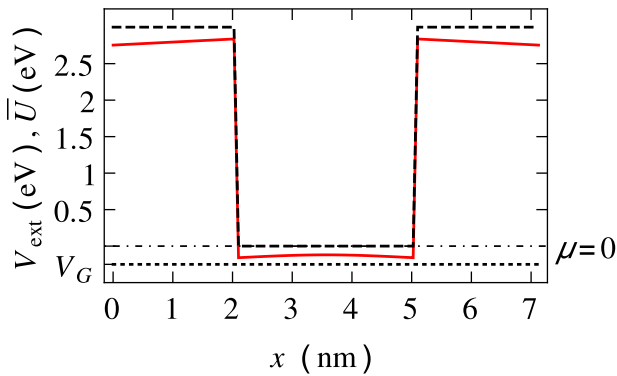}}
\caption{\label{fig:well}
(Color-online)  Potential profile $V_{\textrm{ext}}(x)$ of a 3nm channel GaAs quantum well (black dashed line). We have assumed that the conduction band edge $E_c$ (which is identical to $V_{\textrm{ext}}$) is zero in the channel and at $3 eV$ in the insulator. We use a relative dielectric constant $\epsilon_r=4$ and an effective mass $m = 0.25m_\textrm{e}$, where $m_\textrm{e}$ is the mass of the electron. The chemical potential (dash-dotted line) is $0$ and gate voltage $V_G$ (dotted line) is assumed to be $0.25V$ applied symmetrically. Temperature is assumed to be $300K$. The red solid line corresponds to $\overline{U} = U + V_{\textrm{ext}}$ which is the effective potential energy seen by the electron.}
\end{figure}

The introduction of exchange-correlation effects in GaAs heterostructures introduces only small $(\lesssim$ $\!10\%)$ corrections to the results obtained within the self-consistent Hartree theory and so these effects can be neglected \cite{dassarma1984}.
Thus in GaAsAl$_\textrm{x}$Ga$_{\textrm{1-x}}$As quantum wells where the electrons are confined in one dimension, Eq.~\eqref{eq:spequation} works remarkably well to obtain the electronic potential and the resulting density distribution of the electrons \cite{datta2005quantum}. In \fref{fig:well} we show the external potential $V_{\textrm{ext}}$ seen by the electrons in a 3nm GaAs quantum
well sandwiched between Al$_{\textrm{x}}$Ga$_{\textrm{1-x}}$As barriers.
For such narrow channels only the ground state is significantly occupied (even at room temperature) and  the electronic potential can be calculated by  considering just the ground state and using the local density approximation. 

Working within these approximations, from \eref{eq:spfunctional} and \eref{eq:LMLocal}, we obtain the local SP free-energy functional to be 
\begin{eqnarray}\label{eq:spLDA}
F_{\textrm{LDA}}[V] &=& f^{2D}_x(\lambda_0-\mu) + \frac{1}{8\pi}\int \epsilon |\mathbf{\nabla}V|^{2} dx + S_x \\
&-&\int W_{\textrm{LDA}}[V] \left( -\mathbf{\nabla} \cdot \frac{\epsilon}{4\pi}  \mathbf{\nabla} V  - \rho_{\textrm{LDA}}[V] \right) dx, \nonumber
\end{eqnarray}
where we have suppressed the explicit dependence on $x$ for clarity. In Eq.~\eqref{eq:spLDA},
the density $\rho_{\textrm{LDA}}[V]$ is given by $qf^{2D}_{x}\left|\psi_0(x)\right|^2$ and 
$f_x^{\textrm{2D}}$ is the 2D Fermi function: 
\beq
f^{\textrm{2D}}_x=2t\log\left(1+e^{\beta\left(\lambda_0 - \mu - qV\left(x\right)\right)}\right) 
\eeq
where, $t=ma^2/2\pi\beta\hbar^2$, $m$ is the effective mass of the electron and $a$ is the lattice constant. We note that $f^{\textrm{2D}}_x$ is a scalar function (as oppose to an operator). The factor $t$ comes from integrating the Fermi function over the two dimensions $(yz)$ orthogonal to the direction of confinement. $\lambda_0$ and $\psi_0$ are, respectively, the lowest eigenvalue and eigenfunction of the Schr\"odinger equation
\begin{equation}\label{eq:schrodingerLDA}
\left[-\frac{\hbar^2}{2m}\nabla^2 + V_{\textrm{ext}}(x)\right]\psi(x) = \lambda\psi(x).
\end{equation} 
The function $S_x$, which corresponds to the entropy term of \eref{eq:spfunctional}, is
\beq\label{eq:entropyLDA}
S_x = \frac{t}{\beta}\left[\text{Li}_2\left(1-f_x\right)\right] 
  +\frac{t}{2\beta}  \log^2\left[1-f_x\right]
	+\frac{t}{\beta}\text{Li}_2\left(\frac{f_x-1}{f_x}\right).
\eeq
$S_x$ is obtained after performing the summation over $k_y$ and $k_z$. Here $\textrm{Li}_2[\cdot]$ is polylog function of order $2$ and $f_x$ is the LDA-ground state Fermi function (not an operator) given by
\beq
f_x = \frac{1}{1 + e^{\beta(\lambda_0+qV(x)-\mu)}}.
\eeq
Finally, the function $W_{\textrm{LDA}}[V]$ of \eref{eq:LMLocal} is reduced to 
\begin{eqnarray}\label{eq:LMLDA}
W_{\textrm{LDA}}[V] &=& \rho_{\textrm{LDA}}^{-1}\left(-\mathbf{\nabla} \cdot \frac{\epsilon}{4\pi} \mathbf{\nabla} V\right) \\
&=& \frac{1}{q\beta} \log \left[ \left(e^{-\mathbf{\nabla} \cdot \frac{\epsilon}{4\pi} \mathbf{\nabla} V/\left(2t\left|\psi_0(x)\right|^2\right)}-1\right)e^{\beta(\lambda_0-\mu)}\nonumber\right].
\end{eqnarray}

Minimizing $F_{\textrm{LDA}}$ with respect to $V$ leads to the following relation:
\beq\label{eq:spequationLDA}
\nabla\cdot\left(\frac{\epsilon(x)}{4\pi} \nabla V(x) \right) + qf^{\textrm{2D}}_x\left|\psi_0(x)\right|^2 = 0.
\eeq
This equation can be identified as the SP equation associated with the above example of a quasi-2D quantum well system.

\begin{figure}
\centerline{\includegraphics[scale=0.65]{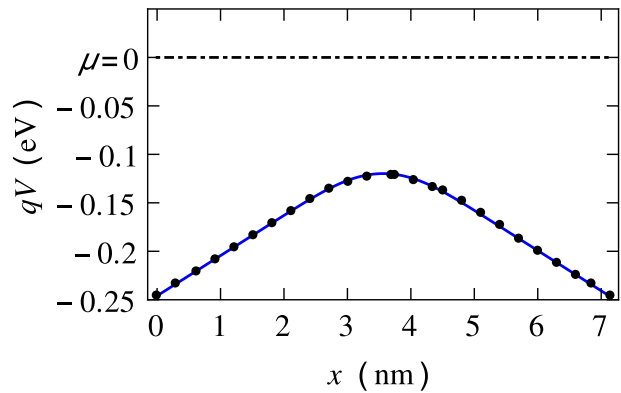}}
\caption{\label{fig:potential}
(Color-online) The variation of the electrostatic potential energy in the quantum well described in \fref{fig:well}. The blue solid curve is obtained from the full Schr\"odinger-Poisson calculation whereas the solid points are obtained from \eref{eq:spequationLDA} which corresponds to the extremum condition of the functional in  \eref{eq:spLDA}. The gate voltage $V_G$ of $0.25V$ provides the boundary conditions for the electrostatic potential.}
\end{figure}

We now demonstrate the validity of the approximations employed to obtain $F_{\textrm{LDA}}$. 
In \fref{fig:potential}, we compare the electrostatic potential energy profile $U(x)=qV(x)$ calculated using both the full SP equation and its aforementioned approximated version, \eref{eq:spequationLDA}.  
The results of the ground-state-LDA calculation are in excellent agreement with the full SP calculation. 
Using this electrostatic potential profile, we compute the effective potential energy seen by the electrons: $\overline{U}=U+V_{\textrm{ext}}$. We plot this effective potential energy $\overline{U}$ (dotted red line) in Fig.~\ref{fig:well}. 

We next consider the opposite regime where the channels are wide enough such that several quantum states are occupied. In this regime the number density of electrons can be well approximated by standard Thomas-Fermi (TF) expression for finite temperatures, in other words, we can assume the electrons to be free and in local thermal equilibrium with the electrostatic potential. Following a process similar to the one employed in arriving at the LDA limit of our local SP functional, we obtain the local free-energy SP functional in the TF limit to be:
\begin{eqnarray}\label{eq:spTF}
F_{\textrm{TF}}[V] &=&  A[V]  + \frac{1}{8\pi}\int \epsilon(x) |\mathbf{\nabla}V(x)|^{2} \nonumber\\
&-&\int W_{\textrm{TF}}[V] \left( -\mathbf{\nabla} \cdot \frac{\epsilon(x)}{4\pi}  \mathbf{\nabla} V (x) - \rho_{\textrm{TF}}[V] \right)dx.\nonumber\\
\end{eqnarray}
In Eq.~\eqref{eq:spTF}, $A[V]$ is the Helmholtz free energy of a free electron gas in 3D. In the computation of $A$, the chemical potential entering the Fermi function is replaced with a position-dependent renormalized form given by $\mu -qV(x) - V_{\textrm{ext}}$. $\rho_{\textrm{TF}}[V] =  2qN_CF_{1/2}(\mu-qV(x)-V_{\textrm{ext}})$ is the Thomas-Fermi density and $W_{\textrm{TF}}[V] = \rho_{\textrm{TF}}^{-1}[-\mathbf{\nabla} \cdot \frac{\epsilon}{4\pi} \mathbf{\nabla}V]$,
where $N_C=(mk_BT/2\pi\hbar^2)^{3/2}$ is the effective density of the conduction band states and
$\rho_{\textrm{TF}}^{-1}$ is the inverse function of $\rho_{\textrm{TF}}$ that can be obtained numerically.

Minimization of $F_{\textrm{TF}}$ with respect to $V$ leads to the condition:
\beq\label{eq:spequationTF}
\nabla\cdot\left(\frac{\epsilon(x)}{4\pi} \nabla V(x) \right) +  2qN_CF_{1/2}(\mu-V(x)-V_{\textrm{ext}})= 0,
\eeq
where $F_{1/2}[\cdot]$ is the Fermi-Dirac integral \cite{Blakemore1982}.  
Eq.~\eqref{eq:spequationTF} is the SP equation for systems treated within the Thomas-Fermi approximation and has been obtained elsewhere \cite{luscombe1990}.
We note that the free-energy functional $F_{\textrm{TF}}$ can be employed to study other quantum nanostructures like quantum dots and quantum wires in the semi-classical regime where the Thomas-Fermi approximation is valid \cite{luscombe1990}. Finally, we note that the application of either approximations, LDA or TF, preserves the properties associated with our general functional.

\section{Conclusion}\label{sec:conclusion}
In this paper, we have derived the true free-energy functional for the Schr\"odinger-Poisson theory. We have obtained the specific forms of our functional for the case where the local-density and ground-state approximations can be applied, and also for the regime where Thomas-Fermi limit holds. Our variational formulation also reveals that functionals, such
as the Nier functional, which are not true free-energy functionals, can be understood as arising from deficient forms of the constraint of SP equation applied to the free energy of the system. In this light, we believe our formulation and the associated free-energy functionals provide a fresh look at the SP theory. On the other hand, our functional can be used for an efficient numerical calculation of the electronic potential in systems obeying Schr\"odinger-Poisson physics as it circumvents the usual approach of solving the coupled Schr\"odinger and Poisson equations iteratively. Furthermore, our functional offers the prospect of on-the-fly optimization under which the SP equation can be solved in conjunction with the update of external parameters.

Schr\"odinger-Poisson theory is equivalent to the self-consistent Hartree theory in quantum many body systems. In this light, our approach towards finding the free-energy functional can be extended to systems beyond Hartree theory. For example, we can apply the same technique in deriving the free-energy functional for the Hartree-Fock systems that have an additional density dependent term in the Schr\"odinger-Poisson equation due to the exchange-correlation potential\cite{lundqvist1971}. 
We hope this paper would lead to further research in this direction.

\section{Acknowledgments}
This work was supported by the Office of the Director of Defense Research and Engineering (DDR$\&$E) and the Air Force Office of Scientific Research (AFOSR) under Award FA9550-10-1-0167. F.J.S acknowledges the support of National Science Foundation (NSF) Grant No. DMR-1309027.


\begin{thebibliography}{32}%
\makeatletter
\providecommand \@ifxundefined [1]{%
 \@ifx{#1\undefined}
}%
\providecommand \@ifnum [1]{%
 \ifnum #1\expandafter \@firstoftwo
 \else \expandafter \@secondoftwo
 \fi
}%
\providecommand \@ifx [1]{%
 \ifx #1\expandafter \@firstoftwo
 \else \expandafter \@secondoftwo
 \fi
}%
\providecommand \natexlab [1]{#1}%
\providecommand \enquote  [1]{``#1''}%
\providecommand \bibnamefont  [1]{#1}%
\providecommand \bibfnamefont [1]{#1}%
\providecommand \citenamefont [1]{#1}%
\providecommand \href@noop [0]{\@secondoftwo}%
\providecommand \href [0]{\begingroup \@sanitize@url \@href}%
\providecommand \@href[1]{\@@startlink{#1}\@@href}%
\providecommand \@@href[1]{\endgroup#1\@@endlink}%
\providecommand \@sanitize@url [0]{\catcode `\\12\catcode `\$12\catcode
  `\&12\catcode `\#12\catcode `\^12\catcode `\_12\catcode `\%12\relax}%
\providecommand \@@startlink[1]{}%
\providecommand \@@endlink[0]{}%
\providecommand \url  [0]{\begingroup\@sanitize@url \@url }%
\providecommand \@url [1]{\endgroup\@href {#1}{\urlprefix }}%
\providecommand \urlprefix  [0]{URL }%
\providecommand \Eprint [0]{\href }%
\providecommand \doibase [0]{http://dx.doi.org/}%
\providecommand \selectlanguage [0]{\@gobble}%
\providecommand \bibinfo  [0]{\@secondoftwo}%
\providecommand \bibfield  [0]{\@secondoftwo}%
\providecommand \translation [1]{[#1]}%
\providecommand \BibitemOpen [0]{}%
\providecommand \bibitemStop [0]{}%
\providecommand \bibitemNoStop [0]{.\EOS\space}%
\providecommand \EOS [0]{\spacefactor3000\relax}%
\providecommand \BibitemShut  [1]{\csname bibitem#1\endcsname}%
\let\auto@bib@innerbib\@empty
\bibitem [{\citenamefont {Ihn}(2010)}]{Ihn}%
  \BibitemOpen
  \bibfield  {author} {\bibinfo {author} {\bibfnamefont {T.}~\bibnamefont
  {Ihn}},\ }\href
  {http://www.amazon.com/Semiconductor-Nanostructures-Quantum-electronic-trans%
port/dp/0199534438} {\emph {\bibinfo {title} {Semiconductor Nanostructures:
  Quantum states and electronic transport}}}\ (\bibinfo  {publisher} {Oxford
  University Press},\ \bibinfo {year} {2010})\BibitemShut {NoStop}%
\bibitem [{\citenamefont {Steiner}(2004)}]{Steiner}%
  \BibitemOpen
  \bibfield  {author} {\bibinfo {author} {\bibfnamefont {T.~D.}\ \bibnamefont
  {Steiner}},\ }\href
  {http://www.amazon.com/Semiconductor-Nanostructures-Optoelectronic-Applicati%
ons-Materials/dp/1580537510/ref=sr_1_3?s=books&ie=UTF8&qid=1378430741&sr=1-3&k%
eywords=semiconductor+nanostructures} {\emph {\bibinfo {title} {Semiconductor
  Nanostructures for Optoelectronic Applications}}}\ (\bibinfo  {publisher}
  {Artech House},\ \bibinfo {year} {2004})\BibitemShut {NoStop}%
\bibitem [{\citenamefont {Nakamura}\ \emph {et~al.}(1996)\citenamefont
  {Nakamura}, \citenamefont {Senoh}, \citenamefont {Nagahama}, \citenamefont
  {Iwasa}, \citenamefont {Yamada}, \citenamefont {Matsushita}, \citenamefont
  {Kiyoku},\ and\ \citenamefont {Sugimoto}}]{nakamura1996ingan}%
  \BibitemOpen
  \bibfield  {author} {\bibinfo {author} {\bibfnamefont {S.}~\bibnamefont
  {Nakamura}}, \bibinfo {author} {\bibfnamefont {M.}~\bibnamefont {Senoh}},
  \bibinfo {author} {\bibfnamefont {S.-i.}\ \bibnamefont {Nagahama}}, \bibinfo
  {author} {\bibfnamefont {N.}~\bibnamefont {Iwasa}}, \bibinfo {author}
  {\bibfnamefont {T.}~\bibnamefont {Yamada}}, \bibinfo {author} {\bibfnamefont
  {T.}~\bibnamefont {Matsushita}}, \bibinfo {author} {\bibfnamefont
  {H.}~\bibnamefont {Kiyoku}}, \ and\ \bibinfo {author} {\bibfnamefont
  {Y.}~\bibnamefont {Sugimoto}},\ }\href@noop {} {\bibfield  {journal}
  {\bibinfo  {journal} {Japanese Journal of Applied Physics-Part 2 Letters}\
  }\textbf {\bibinfo {volume} {35}},\ \bibinfo {pages} {L74} (\bibinfo {year}
  {1996})}\BibitemShut {NoStop}%
\bibitem [{\citenamefont {Tans}\ \emph {et~al.}(1998)\citenamefont {Tans},
  \citenamefont {Verschueren},\ and\ \citenamefont {Dekker}}]{tans1998room}%
  \BibitemOpen
  \bibfield  {author} {\bibinfo {author} {\bibfnamefont {S.~J.}\ \bibnamefont
  {Tans}}, \bibinfo {author} {\bibfnamefont {A.~R.}\ \bibnamefont
  {Verschueren}}, \ and\ \bibinfo {author} {\bibfnamefont {C.}~\bibnamefont
  {Dekker}},\ }\href@noop {} {\bibfield  {journal} {\bibinfo  {journal}
  {Nature}\ }\textbf {\bibinfo {volume} {393}},\ \bibinfo {pages} {49}
  (\bibinfo {year} {1998})}\BibitemShut {NoStop}%
\bibitem [{\citenamefont {Datta}(2005)}]{datta2005quantum}%
  \BibitemOpen
  \bibfield  {author} {\bibinfo {author} {\bibfnamefont {S.}~\bibnamefont
  {Datta}},\ }\href@noop {} {\emph {\bibinfo {title} {Quantum transport: atom
  to transistor}}}\ (\bibinfo  {publisher} {Cambridge University Press},\
  \bibinfo {year} {2005})\BibitemShut {NoStop}%
\bibitem [{\citenamefont {Wettstein}\ \emph {et~al.}(2000)\citenamefont
  {Wettstein}, \citenamefont {Schenk},\ and\ \citenamefont
  {Fichtner}}]{wettstein2000simulation}%
  \BibitemOpen
  \bibfield  {author} {\bibinfo {author} {\bibfnamefont {A.}~\bibnamefont
  {Wettstein}}, \bibinfo {author} {\bibfnamefont {A.}~\bibnamefont {Schenk}}, \
  and\ \bibinfo {author} {\bibfnamefont {W.}~\bibnamefont {Fichtner}},\
  }\href@noop {} {\bibfield  {journal} {\bibinfo  {journal} {IEICE transactions
  on electronics}\ }\textbf {\bibinfo {volume} {83}},\ \bibinfo {pages} {1189}
  (\bibinfo {year} {2000})}\BibitemShut {NoStop}%
\bibitem [{\citenamefont {Kim}\ and\ \citenamefont {Lee}(2000)}]{kim2000self}%
  \BibitemOpen
  \bibfield  {author} {\bibinfo {author} {\bibfnamefont {K.-Y.}\ \bibnamefont
  {Kim}}\ and\ \bibinfo {author} {\bibfnamefont {B.}~\bibnamefont {Lee}},\ }in\
  \href@noop {} {\emph {\bibinfo {booktitle} {Optoelectronic and
  Microelectronic Materials and Devices, 2000. COMMAD 2000. Proceedings
  Conference on}}}\ (\bibinfo {organization} {IEEE},\ \bibinfo {year} {2000})\
  pp.\ \bibinfo {pages} {383--386}\BibitemShut {NoStop}%
\bibitem [{\citenamefont {John}\ \emph {et~al.}(2004)\citenamefont {John},
  \citenamefont {Castro}, \citenamefont {Pereira},\ and\ \citenamefont
  {Pulfrey}}]{john2004schrodinger}%
  \BibitemOpen
  \bibfield  {author} {\bibinfo {author} {\bibfnamefont {D.}~\bibnamefont
  {John}}, \bibinfo {author} {\bibfnamefont {L.}~\bibnamefont {Castro}},
  \bibinfo {author} {\bibfnamefont {P.}~\bibnamefont {Pereira}}, \ and\
  \bibinfo {author} {\bibfnamefont {D.}~\bibnamefont {Pulfrey}},\ }in\
  \href@noop {} {\emph {\bibinfo {booktitle} {Proc. NSTI Nanotech}}},\
  Vol.~\bibinfo {volume} {3}\ (\bibinfo {year} {2004})\ pp.\ \bibinfo {pages}
  {65--68}\BibitemShut {NoStop}%
\bibitem [{\citenamefont {Assad}\ \emph {et~al.}(2000)\citenamefont {Assad},
  \citenamefont {Ren}, \citenamefont {Vasileska}, \citenamefont {Datta},\ and\
  \citenamefont {Lundstrom}}]{assad2000performance}%
  \BibitemOpen
  \bibfield  {author} {\bibinfo {author} {\bibfnamefont {F.}~\bibnamefont
  {Assad}}, \bibinfo {author} {\bibfnamefont {Z.}~\bibnamefont {Ren}}, \bibinfo
  {author} {\bibfnamefont {D.}~\bibnamefont {Vasileska}}, \bibinfo {author}
  {\bibfnamefont {S.}~\bibnamefont {Datta}}, \ and\ \bibinfo {author}
  {\bibfnamefont {M.}~\bibnamefont {Lundstrom}},\ }\href@noop {} {\bibfield
  {journal} {\bibinfo  {journal} {Electron Devices, IEEE Transactions on}\
  }\textbf {\bibinfo {volume} {47}},\ \bibinfo {pages} {232} (\bibinfo {year}
  {2000})}\BibitemShut {NoStop}%
\bibitem [{\citenamefont {Ram-Mohan}\ \emph {et~al.}(2004)\citenamefont
  {Ram-Mohan}, \citenamefont {Yoo},\ and\ \citenamefont
  {Moussa}}]{ram2004schrodinger}%
  \BibitemOpen
  \bibfield  {author} {\bibinfo {author} {\bibfnamefont {L.}~\bibnamefont
  {Ram-Mohan}}, \bibinfo {author} {\bibfnamefont {K.}~\bibnamefont {Yoo}}, \
  and\ \bibinfo {author} {\bibfnamefont {J.}~\bibnamefont {Moussa}},\
  }\href@noop {} {\bibfield  {journal} {\bibinfo  {journal} {Journal of applied
  physics}\ }\textbf {\bibinfo {volume} {95}},\ \bibinfo {pages} {3081}
  (\bibinfo {year} {2004})}\BibitemShut {NoStop}%
\bibitem [{\citenamefont {Ram-Mohan}\ and\ \citenamefont
  {Yoo}(2006)}]{ram2006wavefunction}%
  \BibitemOpen
  \bibfield  {author} {\bibinfo {author} {\bibfnamefont {L.}~\bibnamefont
  {Ram-Mohan}}\ and\ \bibinfo {author} {\bibfnamefont {K.}~\bibnamefont
  {Yoo}},\ }\href@noop {} {\bibfield  {journal} {\bibinfo  {journal} {Journal
  of Physics: Condensed Matter}\ }\textbf {\bibinfo {volume} {18}},\ \bibinfo
  {pages} {R901} (\bibinfo {year} {2006})}\BibitemShut {NoStop}%
\bibitem [{\citenamefont {Pacelli}(1997)}]{pacelli1997self}%
  \BibitemOpen
  \bibfield  {author} {\bibinfo {author} {\bibfnamefont {A.}~\bibnamefont
  {Pacelli}},\ }\href@noop {} {\bibfield  {journal} {\bibinfo  {journal}
  {Electron Devices, IEEE Transactions on}\ }\textbf {\bibinfo {volume} {44}},\
  \bibinfo {pages} {1169} (\bibinfo {year} {1997})}\BibitemShut {NoStop}%
\bibitem [{\citenamefont {Trellakis}\ \emph {et~al.}(1997)\citenamefont
  {Trellakis}, \citenamefont {Galick}, \citenamefont {Pacelli},\ and\
  \citenamefont {Ravaioli}}]{trellakis1997iteration}%
  \BibitemOpen
  \bibfield  {author} {\bibinfo {author} {\bibfnamefont {A.}~\bibnamefont
  {Trellakis}}, \bibinfo {author} {\bibfnamefont {A.}~\bibnamefont {Galick}},
  \bibinfo {author} {\bibfnamefont {A.}~\bibnamefont {Pacelli}}, \ and\
  \bibinfo {author} {\bibfnamefont {U.}~\bibnamefont {Ravaioli}},\ }\href@noop
  {} {\bibfield  {journal} {\bibinfo  {journal} {Journal of Applied Physics}\
  }\textbf {\bibinfo {volume} {81}},\ \bibinfo {pages} {7880} (\bibinfo {year}
  {1997})}\BibitemShut {NoStop}%
\bibitem [{\citenamefont {Tan}\ \emph {et~al.}(1990)\citenamefont {Tan},
  \citenamefont {Snider}, \citenamefont {Chang},\ and\ \citenamefont
  {Hu}}]{tan1990self}%
  \BibitemOpen
  \bibfield  {author} {\bibinfo {author} {\bibfnamefont {I.-H.}\ \bibnamefont
  {Tan}}, \bibinfo {author} {\bibfnamefont {G.}~\bibnamefont {Snider}},
  \bibinfo {author} {\bibfnamefont {L.}~\bibnamefont {Chang}}, \ and\ \bibinfo
  {author} {\bibfnamefont {E.}~\bibnamefont {Hu}},\ }\href@noop {} {\bibfield
  {journal} {\bibinfo  {journal} {Journal of applied physics}\ }\textbf
  {\bibinfo {volume} {68}},\ \bibinfo {pages} {4071} (\bibinfo {year}
  {1990})}\BibitemShut {NoStop}%
\bibitem [{\citenamefont {Luscombe}\ \emph {et~al.}(1992)\citenamefont
  {Luscombe}, \citenamefont {Bouchard},\ and\ \citenamefont
  {Luban}}]{luscombe1992electron}%
  \BibitemOpen
  \bibfield  {author} {\bibinfo {author} {\bibfnamefont {J.~H.}\ \bibnamefont
  {Luscombe}}, \bibinfo {author} {\bibfnamefont {A.~M.}\ \bibnamefont
  {Bouchard}}, \ and\ \bibinfo {author} {\bibfnamefont {M.}~\bibnamefont
  {Luban}},\ }\href@noop {} {\bibfield  {journal} {\bibinfo  {journal}
  {Physical Review B}\ }\textbf {\bibinfo {volume} {46}},\ \bibinfo {pages}
  {10262} (\bibinfo {year} {1992})}\BibitemShut {NoStop}%
\bibitem [{\citenamefont {Car}\ and\ \citenamefont
  {Parrinello}(1985)}]{car-parrinello}%
  \BibitemOpen
  \bibfield  {author} {\bibinfo {author} {\bibfnamefont {R.}~\bibnamefont
  {Car}}\ and\ \bibinfo {author} {\bibfnamefont {M.}~\bibnamefont
  {Parrinello}},\ }\href {\doibase 10.1103/PhysRevLett.55.2471} {\bibfield
  {journal} {\bibinfo  {journal} {Phys. Rev. Lett.}\ }\textbf {\bibinfo
  {volume} {55}},\ \bibinfo {pages} {2471} (\bibinfo {year}
  {1985})}\BibitemShut {NoStop}%
\bibitem [{\citenamefont {Nier}(1993)}]{nier}%
  \BibitemOpen
  \bibfield  {author} {\bibinfo {author} {\bibfnamefont {F.}~\bibnamefont
  {Nier}},\ }\href {\doibase 10.1080/03605309308820966} {\bibfield  {journal}
  {\bibinfo  {journal} {Communications in Partial Differential Equations}\
  }\textbf {\bibinfo {volume} {18}},\ \bibinfo {pages} {1125} (\bibinfo {year}
  {1993})}\BibitemShut {NoStop}%
\bibitem [{\citenamefont {Arriola}\ and\ \citenamefont
  {Soler}(2001)}]{arriola}%
  \BibitemOpen
  \bibfield  {author} {\bibinfo {author} {\bibfnamefont {E.}~\bibnamefont
  {Arriola}}\ and\ \bibinfo {author} {\bibfnamefont {J.}~\bibnamefont
  {Soler}},\ }\href {\doibase 10.1023/A:1010369224196} {\bibfield  {journal}
  {\bibinfo  {journal} {Journal of Statistical Physics}\ }\textbf {\bibinfo
  {volume} {103}},\ \bibinfo {pages} {1069} (\bibinfo {year}
  {2001})}\BibitemShut {NoStop}%
\bibitem [{\citenamefont {Carrillo-Nunez}\ \emph {et~al.}(2010)\citenamefont
  {Carrillo-Nunez}, \citenamefont {Magnus},\ and\ \citenamefont
  {Peeters}}]{carrillo}%
  \BibitemOpen
  \bibfield  {author} {\bibinfo {author} {\bibfnamefont {H.}~\bibnamefont
  {Carrillo-Nunez}}, \bibinfo {author} {\bibfnamefont {W.}~\bibnamefont
  {Magnus}}, \ and\ \bibinfo {author} {\bibfnamefont {F.}~\bibnamefont
  {Peeters}},\ }in\ \href {\doibase 10.1109/SISPAD.2010.5604537} {\emph
  {\bibinfo {booktitle} {Simulation of Semiconductor Processes and Devices
  (SISPAD), 2010 International Conference on}}}\ (\bibinfo {year} {2010})\ pp.\
  \bibinfo {pages} {171--174}\BibitemShut {NoStop}%
\bibitem [{\citenamefont {Solis}\ \emph {et~al.}(2013)\citenamefont {Solis},
  \citenamefont {Jadhao},\ and\ \citenamefont {Olvera de~la Cruz}}]{jso4}%
  \BibitemOpen
  \bibfield  {author} {\bibinfo {author} {\bibfnamefont {F.~J.}\ \bibnamefont
  {Solis}}, \bibinfo {author} {\bibfnamefont {V.}~\bibnamefont {Jadhao}}, \
  and\ \bibinfo {author} {\bibfnamefont {M.}~\bibnamefont {Olvera de~la
  Cruz}},\ }\href {\doibase 10.1103/PhysRevE.88.053306} {\bibfield  {journal}
  {\bibinfo  {journal} {Phys. Rev. E}\ }\textbf {\bibinfo {volume} {88}},\
  \bibinfo {pages} {053306} (\bibinfo {year} {2013})}\BibitemShut {NoStop}%
\bibitem [{\citenamefont {Jadhao}\ \emph {et~al.}(2012)\citenamefont {Jadhao},
  \citenamefont {Solis},\ and\ \citenamefont {Olvera de~la Cruz}}]{jso1}%
  \BibitemOpen
  \bibfield  {author} {\bibinfo {author} {\bibfnamefont {V.}~\bibnamefont
  {Jadhao}}, \bibinfo {author} {\bibfnamefont {F.~J.}\ \bibnamefont {Solis}}, \
  and\ \bibinfo {author} {\bibfnamefont {M.}~\bibnamefont {Olvera de~la
  Cruz}},\ }\href {\doibase 10.1103/PhysRevLett.109.223905} {\bibfield
  {journal} {\bibinfo  {journal} {Phys. Rev. Lett.}\ }\textbf {\bibinfo
  {volume} {109}},\ \bibinfo {pages} {223905} (\bibinfo {year}
  {2012})}\BibitemShut {NoStop}%
\bibitem [{\citenamefont {Jadhao}\ \emph
  {et~al.}(2013{\natexlab{a}})\citenamefont {Jadhao}, \citenamefont {Solis},\
  and\ \citenamefont {Olvera de~la Cruz}}]{jso2}%
  \BibitemOpen
  \bibfield  {author} {\bibinfo {author} {\bibfnamefont {V.}~\bibnamefont
  {Jadhao}}, \bibinfo {author} {\bibfnamefont {F.~J.}\ \bibnamefont {Solis}}, \
  and\ \bibinfo {author} {\bibfnamefont {M.}~\bibnamefont {Olvera de~la
  Cruz}},\ }\href {\doibase 10.1063/1.4789955} {\bibfield  {journal} {\bibinfo
  {journal} {The Journal of Chemical Physics}\ }\textbf {\bibinfo {volume}
  {138}},\ \bibinfo {eid} {054119} (\bibinfo {year}
  {2013}{\natexlab{a}})}\BibitemShut {NoStop}%
\bibitem [{\citenamefont {Jadhao}\ \emph
  {et~al.}(2013{\natexlab{b}})\citenamefont {Jadhao}, \citenamefont {Solis},\
  and\ \citenamefont {Olvera de~la Cruz}}]{jso3}%
  \BibitemOpen
  \bibfield  {author} {\bibinfo {author} {\bibfnamefont {V.}~\bibnamefont
  {Jadhao}}, \bibinfo {author} {\bibfnamefont {F.~J.}\ \bibnamefont {Solis}}, \
  and\ \bibinfo {author} {\bibfnamefont {M.}~\bibnamefont {Olvera de~la
  Cruz}},\ }\href {\doibase 10.1103/PhysRevE.88.022305} {\bibfield  {journal}
  {\bibinfo  {journal} {Phys. Rev. E}\ }\textbf {\bibinfo {volume} {88}},\
  \bibinfo {pages} {022305} (\bibinfo {year} {2013}{\natexlab{b}})}\BibitemShut
  {NoStop}%
\bibitem [{Note1()}]{Note1}%
  \BibitemOpen
  \bibinfo {note} {See Eq.~(33) of Ref. 20}\BibitemShut {NoStop}%
\bibitem [{Note2()}]{Note2}%
  \BibitemOpen
  \bibinfo {note} {See Eq.~(41) of Ref. 20}\BibitemShut {NoStop}%
\bibitem [{\citenamefont {Stern}\ and\ \citenamefont
  {Das~Sarma}(1984)}]{dassarma1984}%
  \BibitemOpen
  \bibfield  {author} {\bibinfo {author} {\bibfnamefont {F.}~\bibnamefont
  {Stern}}\ and\ \bibinfo {author} {\bibfnamefont {S.}~\bibnamefont
  {Das~Sarma}},\ }\href {\doibase 10.1103/PhysRevB.30.840} {\bibfield
  {journal} {\bibinfo  {journal} {Phys. Rev. B}\ }\textbf {\bibinfo {volume}
  {30}},\ \bibinfo {pages} {840} (\bibinfo {year} {1984})}\BibitemShut
  {NoStop}%
\bibitem [{\citenamefont {Ortalano}\ \emph {et~al.}(1997)\citenamefont
  {Ortalano}, \citenamefont {He},\ and\ \citenamefont
  {Das~Sarma}}]{dassarma1997}%
  \BibitemOpen
  \bibfield  {author} {\bibinfo {author} {\bibfnamefont {M.~W.}\ \bibnamefont
  {Ortalano}}, \bibinfo {author} {\bibfnamefont {S.}~\bibnamefont {He}}, \ and\
  \bibinfo {author} {\bibfnamefont {S.}~\bibnamefont {Das~Sarma}},\ }\href
  {\doibase 10.1103/PhysRevB.55.7702} {\bibfield  {journal} {\bibinfo
  {journal} {Phys. Rev. B}\ }\textbf {\bibinfo {volume} {55}},\ \bibinfo
  {pages} {7702} (\bibinfo {year} {1997})}\BibitemShut {NoStop}%
\bibitem [{\citenamefont {Schmerek}\ and\ \citenamefont
  {Hansen}(1999)}]{hansen1999}%
  \BibitemOpen
  \bibfield  {author} {\bibinfo {author} {\bibfnamefont {D.}~\bibnamefont
  {Schmerek}}\ and\ \bibinfo {author} {\bibfnamefont {W.}~\bibnamefont
  {Hansen}},\ }\href {\doibase 10.1103/PhysRevB.60.4485} {\bibfield  {journal}
  {\bibinfo  {journal} {Phys. Rev. B}\ }\textbf {\bibinfo {volume} {60}},\
  \bibinfo {pages} {4485} (\bibinfo {year} {1999})}\BibitemShut {NoStop}%
\bibitem [{\citenamefont {Nagaraja}\ \emph {et~al.}(1997)\citenamefont
  {Nagaraja}, \citenamefont {Matagne}, \citenamefont {Thean}, \citenamefont
  {Leburton}, \citenamefont {Kim},\ and\ \citenamefont {Martin}}]{nagraja1997}%
  \BibitemOpen
  \bibfield  {author} {\bibinfo {author} {\bibfnamefont {S.}~\bibnamefont
  {Nagaraja}}, \bibinfo {author} {\bibfnamefont {P.}~\bibnamefont {Matagne}},
  \bibinfo {author} {\bibfnamefont {V.-Y.}\ \bibnamefont {Thean}}, \bibinfo
  {author} {\bibfnamefont {J.-P.}\ \bibnamefont {Leburton}}, \bibinfo {author}
  {\bibfnamefont {Y.-H.}\ \bibnamefont {Kim}}, \ and\ \bibinfo {author}
  {\bibfnamefont {R.~M.}\ \bibnamefont {Martin}},\ }\href {\doibase
  10.1103/PhysRevB.56.15752} {\bibfield  {journal} {\bibinfo  {journal} {Phys.
  Rev. B}\ }\textbf {\bibinfo {volume} {56}},\ \bibinfo {pages} {15752}
  (\bibinfo {year} {1997})}\BibitemShut {NoStop}%
\bibitem [{\citenamefont {Hedin}\ and\ \citenamefont
  {Lundqvist}(1971)}]{lundqvist1971}%
  \BibitemOpen
  \bibfield  {author} {\bibinfo {author} {\bibfnamefont {L.}~\bibnamefont
  {Hedin}}\ and\ \bibinfo {author} {\bibfnamefont {B.~I.}\ \bibnamefont
  {Lundqvist}},\ }\href {http://stacks.iop.org/0022-3719/4/i=14/a=022}
  {\bibfield  {journal} {\bibinfo  {journal} {Journal of Physics C: Solid State
  Physics}\ }\textbf {\bibinfo {volume} {4}},\ \bibinfo {pages} {2064}
  (\bibinfo {year} {1971})}\BibitemShut {NoStop}%
\bibitem [{\citenamefont {Blakemore}(1982)}]{Blakemore1982}%
  \BibitemOpen
  \bibfield  {author} {\bibinfo {author} {\bibfnamefont {J.}~\bibnamefont
  {Blakemore}},\ }\href {\doibase
  http://dx.doi.org/10.1016/0038-1101(82)90143-5} {\bibfield  {journal}
  {\bibinfo  {journal} {Solid-State Electronics}\ }\textbf {\bibinfo {volume}
  {25}},\ \bibinfo {pages} {1067 } (\bibinfo {year} {1982})}\BibitemShut
  {NoStop}%
\bibitem [{\citenamefont {Luscombe}\ and\ \citenamefont
  {Luban}(1990)}]{luscombe1990}%
  \BibitemOpen
  \bibfield  {author} {\bibinfo {author} {\bibfnamefont {J.~H.}\ \bibnamefont
  {Luscombe}}\ and\ \bibinfo {author} {\bibfnamefont {M.}~\bibnamefont
  {Luban}},\ }\href {\doibase http://dx.doi.org/10.1063/1.103578} {\bibfield
  {journal} {\bibinfo  {journal} {Applied Physics Letters}\ }\textbf {\bibinfo
  {volume} {57}},\ \bibinfo {pages} {61} (\bibinfo {year} {1990})}\BibitemShut
  {NoStop}%
\end{thebibliography}
%

\end{document}